\documentclass[english, preprint,12pt]{elsarticle}

\usepackage{amssymb}
\usepackage{array}
\usepackage{units}
\usepackage{graphicx}

\usepackage{mathrsfs}

\usepackage{epsfig}
\usepackage{natbib}

\usepackage{morefloats}

\makeatother

\usepackage{babel}

\makeatother

\usepackage{babel}
\usepackage{color}

\begin{document}

\begin{frontmatter}

\title{Spatial and temporal structures of four financial markets in Greater China}

 \author{F. Y. Ouyang}
 \author{B. Zheng\corref{cor1}}
 \author{X. F. Jiang}

 \address{Department of Physics, Zhejiang University, Hangzhou 310027, P. R. China}
 \cortext[cor1]{Corresponding author. Email address: zheng@zimp.zju.edu.cn}

\begin{abstract}

We investigate the spatial and temporal structures of four financial markets in Greater China. In particular, we uncover different characteristics of the four markets by analyzing the sector and subsector structures which are detected through the random matrix theory. Meanwhile, we observe that the Taiwan and Hongkong stock markets show a negative return-volatility correlation, i.e., the so-called leverage effect. The Shanghai and Shenzhen stock markets are more complicated. Before the year 2000, the two markets exhibit a strong positive return-volatility correlation, which is called the anti-leverage effect. After 2000, however, it gradually changes to the leverage effect. We also find that the recurrence interval distributions of both the trading volume volatilities and price volatilities follow a power law behavior, while the exponents vary among different markets.
\end{abstract}


\end{frontmatter}

\section{Introduction}
Financial markets are complex systems with many-body interactions. In recent years, much attention of physicists has been paid to the financial dynamics, and physical concepts and methods are applied to analyze the dynamic behavior. As large amounts of financial data are available now, it allows to explore the fine structure of the financial dynamics and achieve various empirical results~\cite{man95,ple99a,gop99,gia01,bou01,she09a,pod09,pod10}. With rapid development of the economy, the financial markets in Greater China attract more attention from the world. Let us now focus on four stock markets, i.e., the Shanghai stock market, Shenzhen stock market, Taiwan stock market and Hongkong stock market.
Due to different political and economic systems, the dynamic behavior varies much among the four markets. The economy style of Taiwan is a typical export-oriented one. The stock market developed much through several important economic policies, such as import substitution, export expansion and structural adjustment.
Hongkong is a financial center in Asia, and the economy is prosperous.
The Shanghai and Shenzhen stock markets are both in mainland China, and undergoing a rapid development in recent years. To the best of our knowledge, there have not been literatures focusing on the comparative study of the spatial and temporal structures of the four stock markets, although some relevant works could be found such as the comparison between the response dynamics in transition economies and developed countries~\cite{ten10}.
In this paper, we intend to provide a comparative study about the four stock markets, and understand
how political and economic environments may influence the financial dynamics.

In the past years the properties of the cross-correlation matrix of individual stock prices have been analyzed, e.g.,
with the random matrix theory (RMT), and much effort has been made to identify the business sectors
by the components in the eigenvectors of the cross-correlation matrix~\cite{lal99,ple99,ple02a,uts04,pan07a,gar08,she09,oh11}. In this paper, the analysis of the so-called spatial structure is just an analysis about the cross-correlations between individual stocks based on the RMT theory.
After taking into account the signs of the components in an eigenvector, a sector may be further separated into two subsectors, i.e., the positive and negative
subsectors~\cite{jia12}.
A purpose of this paper is to investigate the spatial structures of the four stock markets in Greater China, and uncover characteristics of the sector and subsector structures for each market.

The dynamic behavior of the stock prices
has been studied for years, and various results have been obtained. For example, the probability distribution of the price return usually exhibits a power-law tail, the price volatility is long-range correlated in time, while the price return itself is short-range correlated~\cite{ple99a,lux96,pan07,li11}.
To better understand the dynamic behavior of the stock prices,
one may consider a
higher-order time correlation, i.e., the return-volatility correlation~\cite{bou01,she09a,pan07a,qiu06}. A negative return-volatility correlation, which is called the leverage effect, was first discovered by Black in 1976~\cite{bla76,cox76}.
The leverage effect is observed in most of the stock markets in the world, while a positive return-volatility correlation,
which is called the anti-leverage effect, was detected in the stock markets of mainland China~\cite{man95,gia01,she09a}. The leverage and anti-leverage effects are crucial for the understanding of the price dynamics~\cite{bou01,she09a,qiu06}.
In this study, we analyze the return-volatility correlation function of the four corresponding stock-market indices, i.e., the  Shanghai Composite Index, Shenzhen Composite Index, Taiwan Weighted Index and Hang Seng Index.

The analysis of the recurrence interval may deepen the understanding of the dynamic behavior in financial markets~\cite{li11,ren10}.
Recently, statistical properties of the recurrence intervals of volume volatilities and price volatilities have been studied~\cite{ren08,ren09,ren09a,qiu08}.
We present a comparative study on the recurrence interval distributions of the four stock markets. For each market, we analyze the recurrence interval distributions for both the trading volume volatilities and the price volatilities.

The paper is organized as follows.
In section 2, we investigate the sector and subsector
structures. In section 3, we analyze the return-volatility correlation function
and the distributions of the recurrence intervals for both volume volatilities and price volatilities.
In section 4, we present the conclusion.

\section{Sector and subsector structures}

We define the logarithmic price return of the $i$-th stock over a
time interval $\Delta t$ as \begin{equation}
R_{i}(t',\triangle t)\equiv ln\, P_{i}(t'+\triangle t)-ln\, P_{i}(t'),\end{equation}
where $P_{i}(t')$ represents the close price at time $t'$, and we set $\Delta t$ to be one day.
To ensure that the results are independent of the fluctuation scales of the stock prices,
we introduce the normalized return of the $i$ -th stock
\begin{equation}
r_{i}(t',\triangle t)=\frac{R_{i}-\langle \ensuremath{R_{i}}\rangle }{\sigma_{i}},
\end{equation}
where $\langle \cdots\rangle$ represents the time average over time $t'$ and
the standard deviation of $R_{i}$ is denoted by $\sigma_{i}=\sqrt{\langle R_{i}^{2}\rangle -\langle R_{i}\rangle ^{2}}$~\cite{jia13}. Then, the elements of the equal-time
cross-correlation matrix $C$ are defined by \begin{equation}
C_{ij}\equiv \langle r_{i}(t')r_{j}(t')\rangle ,\end{equation}
which measure the correlations between the returns of individual stocks. According to the definition, $C$ is a real symmetric matrix
with $C_{ii}=1.$ The value of $C_{ij}$ ranges from $-1$ to $1$.

\begin{figure}[ht]
\begin{center}
\includegraphics[scale=0.35]{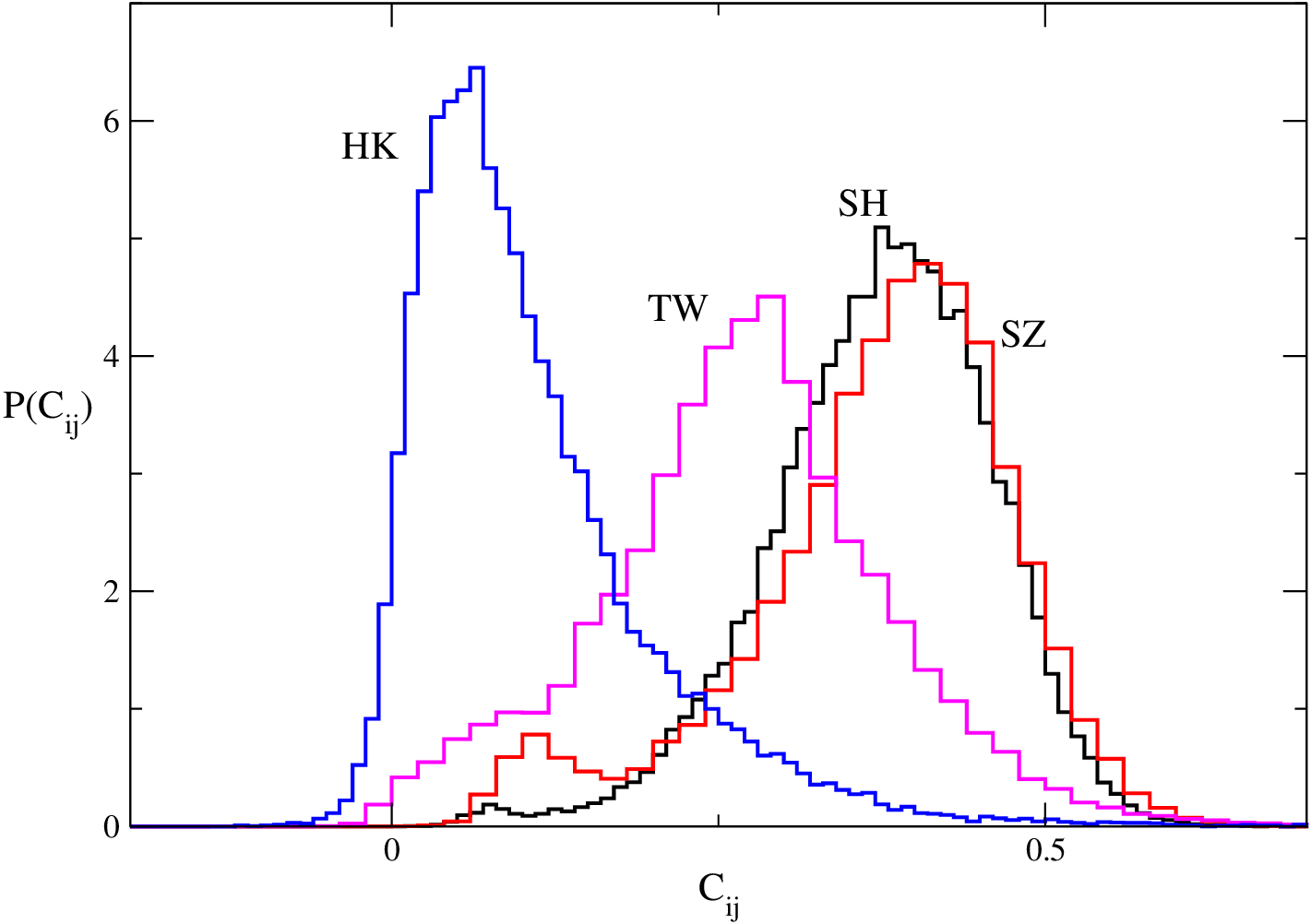}

\caption{The probability distributions of $C_{ij}$ for the SH, SZ, TW and HK stock markets.
The abbreviations of the stock markets are introduced in the caption of Table~\ref{t0}.}
{\footnotesize \label{f1}}
\end{center}
\end{figure}

In this paper, we compute the cross-correlation matrix $C$ with the daily stock prices of the four stock markets in Greater China.
The time periods of 259 stocks for each stock market are shown in the first column of Table~\ref{t0}. Why do we choose 259 stocks for each stock market? On the one hand, we should use as many stocks as we can. On the other hand, the available data of the stocks should be as long as possible. Under these conditions we obtain 259 stocks for the Shanghai stock market. For comparison, we also use this number 259 for the other three markets. The probability distributions $P(C_{ij})$
of the four stock markets are displayed in Fig.~\ref{f1}. The average value of $C_{ij}$ is close to 0.37 for both the Shanghai and Shenzhen markets,
larger than 0.26 for the Taiwan market, and much larger than 0.11 for the Hongkong market.
Previous studies show that the average value of $C_{ij}$ of a mature market is smaller than that of an emerging market~\cite{mor00,pod06}. Our results indicate that the Hongkong stock market is a mature one, the Shanghai and Shenzhen stock markets are emerging, and the Taiwan stock market is in between.

We then compute the eigenvalues of the cross-correlation  matrix $C$,
and compare it with the so-called Wishart matrix~\cite{dys71,sen99}. The Wishart matrix is
derived from non-correlated time series. Assuming $N$ time series
with length $T$, statistical properties of such random matrices are
known. In the limit $N$$\rightarrow\infty$ and $T$$\rightarrow\infty$
with $Q\equiv T/N$$\geqslant1$, the probability distribution $P_{rm}(\lambda)$
of the eigenvalue $\lambda$ can be given by~\cite{dys71,sen99}\begin{equation}
P_{rm}(\lambda)=\frac{Q}{2\pi}\frac{\sqrt{(\lambda_{max}^{ran}-\lambda)(\lambda-\lambda_{min}^{ran})}}{\lambda},\end{equation}
the lower and upper bounds are\begin{equation}
\lambda_{min(max)}^{ran}=\left[1\pm(1/\sqrt{Q})\right]^{2},\end{equation}
where $\lambda_{min}^{ran}\leq\lambda\leq\lambda_{max}^{ran}$.

\begin{figure}[ht]
\begin{center}
{\footnotesize \includegraphics[scale=0.35]{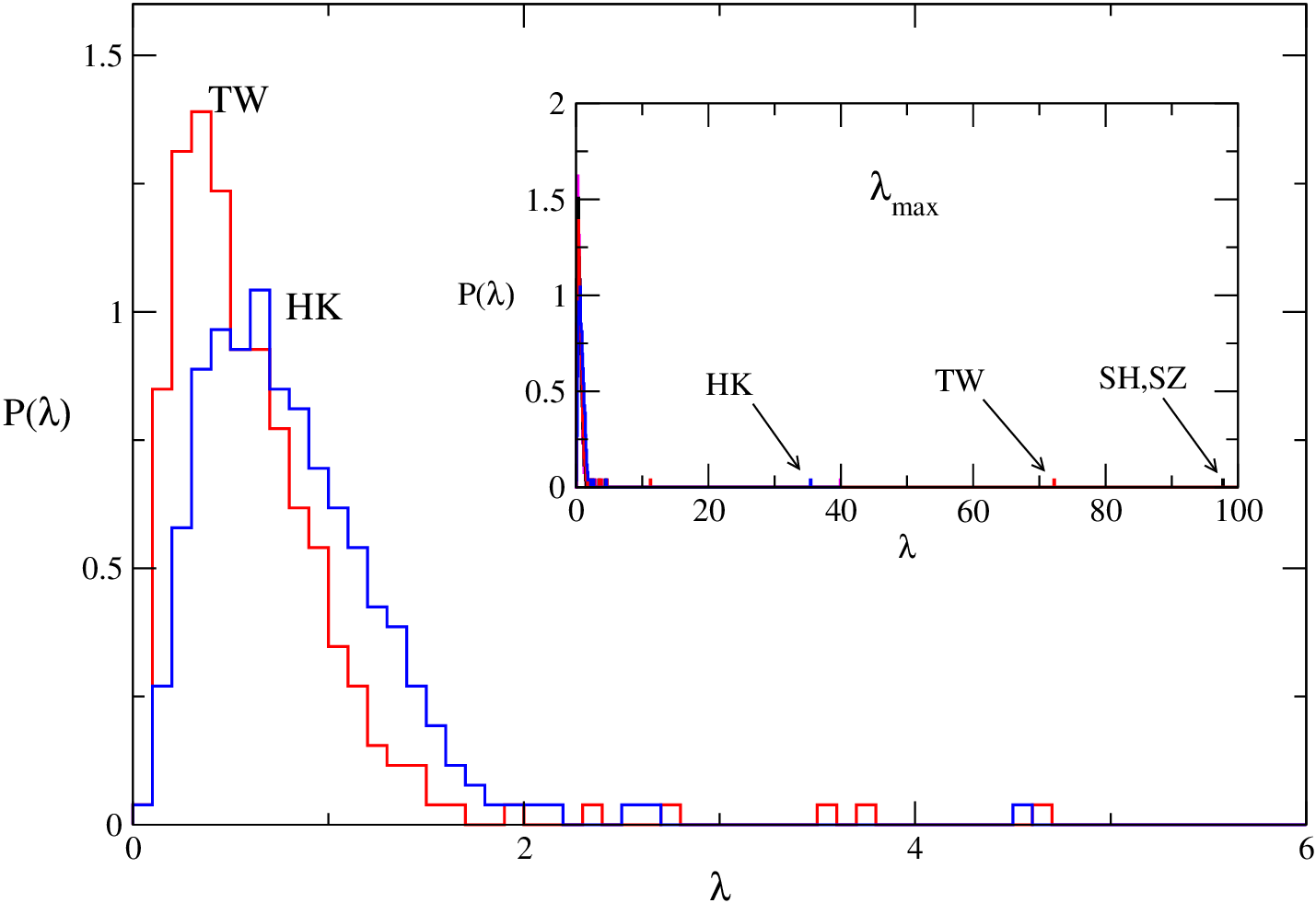}}{\footnotesize \par}

\caption{The probability distributions of the eigenvalues of the correlation
matrix $C$ for the TW and HK stock markets. The inset shows the largest eigenvalue for each market. $\lambda_{max}$ for the SH and SZ stock markets are about 98.0, while those for the TW and HK stock markets are 72.3 and 35.5 respectively.}

{\footnotesize \label{f2}}
\end{center}
\end{figure}

For a dynamic system, large eigenvalues deviate from $P_{rm}(\lambda)$, implying that there exist non-random
interactions. In fact, both mature and emerging stock markets show
the same phenomenon that the bulk of the eigenvalue spectrum $P(\lambda)$
of the cross-correlation matrix is similar to $P_{rm}(\lambda)$ of
the Wishart matrix~\cite{ple02a,pan07a,she09,gop01}, but some large eigenvalues deviate greatly from
the upper bound $\lambda_{max}^{ran}$. The probability distributions $P(\lambda)$ of the Taiwan and Hongkong stock markets are shown in Fig.~\ref{f2}. $P(\lambda)$ of the Shanghai and Shenzhen markets are similar, therefore are not shown in the figure.
The inset displays $\lambda_{max}$ for each stock market.

For our datasets, N is equal to 259 for all four stock markets, T is about 2000 days. As shown in Table~\ref{t0}, $\lambda_{max}^{ran}$ takes similar values for the four markets, according to Eq.(5). $\lambda_{max}$ is rather stable for sufficiently large T. The dependence of $\lambda_{max}$ on T is already stable for T around 2000. Therefore, the comparison of $\lambda_{max}$ is meaningful. $\lambda_{max}$ for the Shanghai and Shenzhen markets are about 98.0,
while the ones for the Taiwan and Hongkong markets are 72.3 and 35.5 respectively. It is known that $\lambda_{max}$
for the US stock market is about 40.0~\cite{she09}. $\lambda_{max}$ for the Hongkong market is close to the one for the US market,
while $\lambda_{max}$ for the Shanghai, Shenzhen and Taiwan stock markets are much larger. It indicates that the Hongkong stock market is
mature, similar to the western stock markets, while the Shanghai and Shenzhen stock markets are emerging. The Taiwan stock market is
in between.

The eigenvector of a large eigenvalue is dominated by a group of stocks, usually associated
with a business sector~\cite{she09,bap02,man00,bou03}. Let $u_{i}(\lambda_{\alpha})$ denotes the component of the $i$-th stock in the eigenvector of $\lambda_{\alpha}$.
In order to identify the sector structure, we introduce a threshold $u_{c}$: $|u_{i}(\lambda_{\alpha})|\geq u_{c}$, to select the dominating components in a particular eigenvector. The threshold $u_{c}$ should be properly determined. Firstly, for a random matrix, $\langle |u(\lambda)|\rangle $$\sim1/\sqrt{N}$ for every eigenmode, where $N$
is the number of stocks. Therefore,
the threshold should be larger than this value. Secondly, $u_{c}$ should
not be too large, because for a large threshold there are not so many stocks in each sector~\cite{jia12}. We choose different thresholds to test the stability of the results, such as $u_{c}=0.06,0.08,0.10,0.12$. In fact, the results are stable when changing the value of the threshold. We just show the typical ones in our paper.
After taking into account the signs of the components in an eigenvector, we may separate a sector into two subsectors by introducing two thresholds $u_{c}^{\pm}$$=\pm u_{c}$: $u_{i}$$(\lambda_{\alpha})\geq u_{c}^{+}$ and $u_{i}$$(\lambda_{\alpha})$$\leq u_{c}^{-}$, which correspond to the positive and negative subsectors respectively~\cite{jia12}.

With the above method, we identify the sector and subsector structures up to the ninth largest eigenvalue $\lambda_{8}$. The largest eigenvalue $\lambda_{0}$ represents the market mode, which is driven by interactions common for stocks in the entire market~\cite{jia12}.
Therefore the market mode does not correspond to a sector.
If the financial situations of a company in the Shanghai market are abnormal, the company will be treated specially, and a prefix of the acronym {}``ST'' will
be added to the stock ticker. The abnormal financial situations include: the audited profits are negative in two successive accounting years, the audited net worth per share is less than its stock's par value in the recent accounting year. The acronym {}``ST'' will be removed
when the financial situations become normal~\cite{she09}.
The positive components in the eigenvector of the second largest eigenvalue $\lambda_{1}$ for the Shanghai market are dominated by the ST stocks. Therefore, we define this subsector as the ST subsector. Since more than half of the stocks in the negative components in the eigenvector
of $\lambda_{1}$ belong
to industrial goods, we call it the Industrial goods subsector.
The positive and negative subsectors of $\lambda_{2}$ are identified to be Real estate
and Health care subsectors. The detected subsectors of other eigenvalues are displayed in Table~\ref{t1}, while {}``Null'' represents the one we could not identify. The subsectors for the Shenzhen, Taiwan and Hongkong stock markets are shown in Table~\ref{t2}. From Table~\ref{t1} and~\ref{t2}, we can see that a particular subsector may appear several times for a stock market, and the one which shows up the most often plays a dominating role in this market. For the Shanghai stock market, there are fourteen identified subsectors, and the dominating ones are Basic materials and Industrial goods.
For the Shenzhen stock market, the dominating ones are the Energy and Real estate subsectors.

For the Taiwan stock market, the negative components in the eigenvector of
$\lambda_{1}$ are dominated by the electronic industry stocks which defines the Electronic industry subsector, while the positive ones remain unknown. The negative components of $\lambda_{3}$ and the positive ones of $\lambda_{4}$ are identified as the Steel industry subsectors.
The dominating subsectors of this market are the Chemical industry and Daily consumer
goods.

For the Hongkong stock market, the finance and real estate stocks are strongly correlated with
each other.
The negative components of $\lambda_{3}$ are dominated by the finance and real estate stocks which defines the Real estate \& Finance subsector. The positive components of $\lambda_{3}$ can be identified as the Service subsector.
The dominating subsector of the Hongkong stock market is the
Real estate \& Finance.

To better understand the above results, let us look at the regional economies of the four stock markets. Shanghai and Shenzhen are both in mainland China where the business such as basic material, energy and industry plays a vital role ~\cite{roy01,yan08a,hui00}. The real estate attracts much attention in Shenzhen, and it also plays an important role in the economy of mainland China~\cite{sin00}. The economy of Taiwan is forefront in Asia. In Taiwan, there are various types of industries, and the dominating businesses are steel, machinery, computer and electronics, textile and clothing ~\cite{fee99}. Hongkong is changing from a port city into an industrial one. In Hongkong, the public service is very important, the businesses such as
real estate, finance, commerce, tourism and some other traditional industries are dominating~\cite{cha01c}.
Our results of the sector and subsector structure in Tables~\ref{t1} and \ref{t2} well reflect the features of the regional economies.

Now we investigate the anti-correlation between the two subsectors for each eigenmode. The cross-correlation between two stocks can be decomposed
into different eigenmodes,  \begin{equation}\label{eq2}
C_{ij}=\sum_{\alpha=1}^{N}\lambda_{\alpha}C_{ij}^{\alpha}, \quad C_{ij}^{\alpha}=u_{i}^{\alpha}u_{j}^{\alpha},\end{equation} where $\lambda_{\alpha}$ is the $\alpha$-th eigenvalue, $u_{i}^{\alpha}$
is the $i$-th component in the eigenvector of $\lambda_{\alpha}$,
and $C_{ij}^{\alpha}$ represents the cross-correlation in the $\alpha$-th
eigenmode. The eigenvalue $\lambda_{\alpha}$ is always positive, and it gives the weight to the $\alpha$-th eigenmode.
According to Eq.(\ref{eq2}), $C_{ij}^{\alpha}$
is positive when the components $u_{i}^{\alpha}$ and $u_{j}^{\alpha}$
have a same sign. Otherwise, it is negative. When $C_{ij}^{\alpha}$
is negative, two stocks are referred to be anti-correlated in
this eigenmode, indicating that when the price of the $i$-th stock
rises, the price of the $j$-th stock tends to fall, and vice versa~\cite{jia12}. Therefore, any two stocks in a particular subsector are positively correlated in this eigenmode, while those in different subsectors are anti-correlated.

\begin{figure}[ht]
\begin{center}
\includegraphics[scale=0.35]{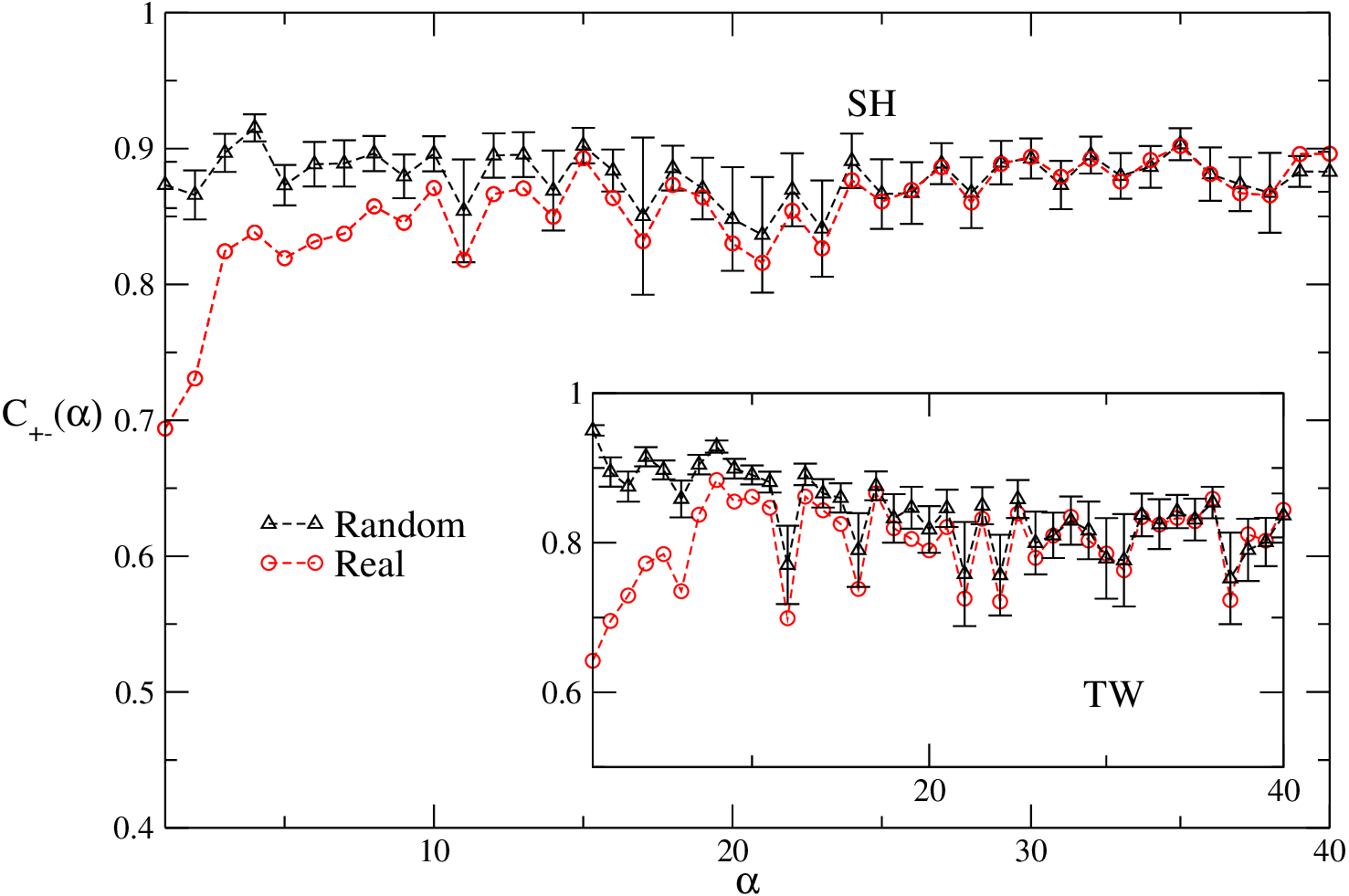}\caption{$C_{+-}(\alpha)$ for the SH and TW stock markets compared with that between two random combinations of stock price returns. Error bars are given to the random curves.}

{\footnotesize \label{f4}}
\end{center}
\end{figure}

To measure the anti-correlation between the positive and negative subsectors quantitatively, we construct the combinations of stock price returns
in the two subsectors, $I_{\alpha}^{\pm}(t)=\sum_{i}u_{i}^{\pm}(\alpha)r_{i}(t),$
and compute the cross-correlation between $I_{\alpha}^{+}(t)$ and $I_{\alpha}^{-}(t)$ as
\begin{equation}
C_{+-}(\alpha)=\langle I_{\alpha}^{+}(t)I_{\alpha}^{-}(t)\rangle,\end{equation}
where $u_{i}^{\pm}(\alpha)$ is the $i$-th positive or negative component
in the $\alpha$-th eigenmode selected by a threshold~\cite{jia12}. $u_{c}$ could be different from this market to another market since different markets may have different fluctuation scales.
We choose $u_{c}^{\pm}=\pm0.06$
for the Taiwan stock market, and $u_{c}^{\pm}=\pm0.08$ for the Shanghai, Shenzhen and Hongkong stock markets.
The cross-correlation $C_{+-}(\alpha)$, in comparison with that between two random combinations are shown in Fig.\,\ref{f4}. For each market, $C_{+-}(\alpha)$ increases slowly with increasing $\alpha$, and gradually approaches the curve calculated from two random combinations of stock price returns.

\begin{figure}[t]
\begin{center}
{\footnotesize \includegraphics[scale=0.35]{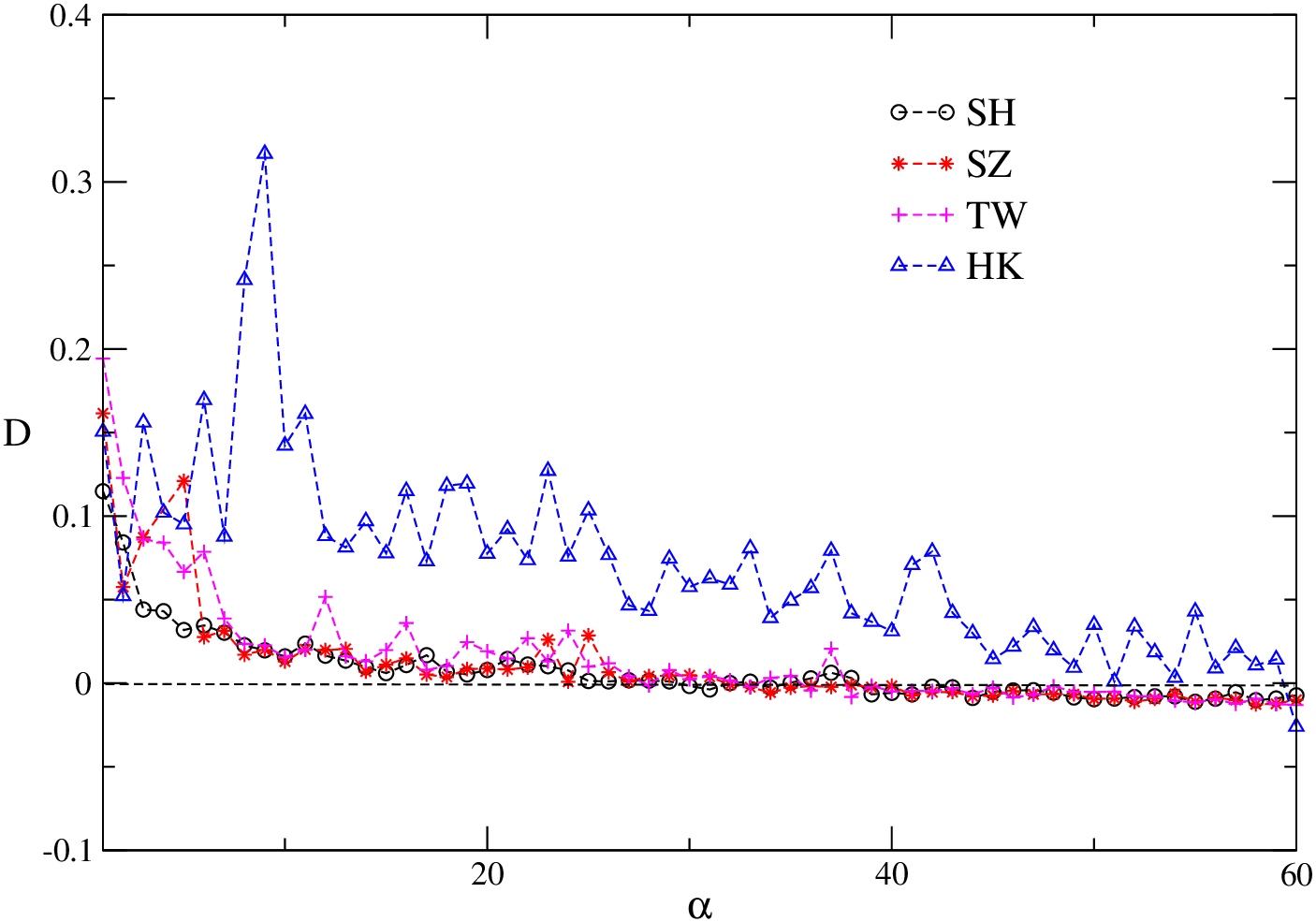}}{\footnotesize \par}

\caption{The quantity $D$ for each stock market.}

{\footnotesize \label{f7}}
\end{center}
\end{figure}

We denote the cross-correlation $C_{+-}(\alpha)$ computed with $I_{\alpha}^{+}(t)$ and $I_{\alpha}^{-}(t)$ as $C_{real}$, and the one computed with two random combinations of stock price returns as $C_{rand}$. The relative difference $D$ between $C_{real}$ and $C_{rand}$ can be defined as
\begin{equation}
D=(C_{rand}-C_{real})/(C_{rand}+C_{real}).
\end{equation}
The results for the four stock markets are shown in Fig.\,\ref{f7}. A positive value of $D$
indicates the existence of the anti-correlation between the positive and negative subsectors. For the Shanghai, Shenzhen and Taiwan markets, $D$ drops to zero for $\alpha \approx 20$. While for the Hongkong market, the value of $D$ is larger, and it remains non-zero up to $\alpha \approx 40$, which implies that the anti-correlation between positive and negative subsectors of the Hongkong stock market is much stronger.

Why would there be the subsector structure with the anti-correlation between the positive and negative subsectors?
For the Shanghai and Shenzhen stock markets, the positive and negative subsectors of $\lambda_{2}$ are Real estate and Health care respectively.
It is well known that the real estate in mainland China fluctuates much with the macroeconomy. When the economy is prosperous, the real estate will also be booming, while the health care will be less dependent on the economy. For the Taiwan market, we take the two subsectors of $\lambda_{3}$ as an example. As described in Ref.~\cite{jia12}, from the intrinsic properties, the steel industry is classified as the strongly cyclical industry, while the daily consumer goods is classified as the weakly cyclical industry. The weakly and strongly cyclical industries are weakly and strongly correlated with the macro-economy environment, respectively. Thus, the strongly cyclical industry is preferred when the macro-economy is booming. However, the investors would rather choose the weakly cyclical industry when the macro-economy declines. The result in Ref.~\cite{wor97} also shows that the energy such as the steel industry fluctuates much with the economy.
For the Hongkong market, the two subsectors of $\lambda_{3}$ are Real estate \& Finance and Service. As we have described above, the real estate \& finance fluctuate with the economy, while the service will remain relatively stable.

\section{Dynamic properties}

We first consider the return-volatility correlation function defined as\begin{equation}
L(t)=[\langle r(t')|r(t'+t)|^{2} \rangle-L_{0}]/Z,\end{equation}
with $Z=\langle |r(t')|^{2} \rangle ^{2}$ and $L_{0}=\langle r(t') \rangle \langle |r(t')|^{2} \rangle$. For $t > 0$, $L(t)$ describes how the past returns affect
the future volatilities. It was first observed by Black that the past negative returns increase future volatilities,
i.e., the return-volatility correlation is negative~\cite{bla76,cox76}, and this phenomenon is called the leverage effect.
The leverage effect is observed in most of the stock markets in the world.
However, a positive return-volatility correlation, which is now called the anti-leverage effect, is observed in the stock markets in mainland China,
and this is the discovery a few years ago~\cite{she09a,qiu06,qiu07}.

\begin{figure}[ht]
\begin{center}
{\footnotesize \includegraphics[scale=0.35]{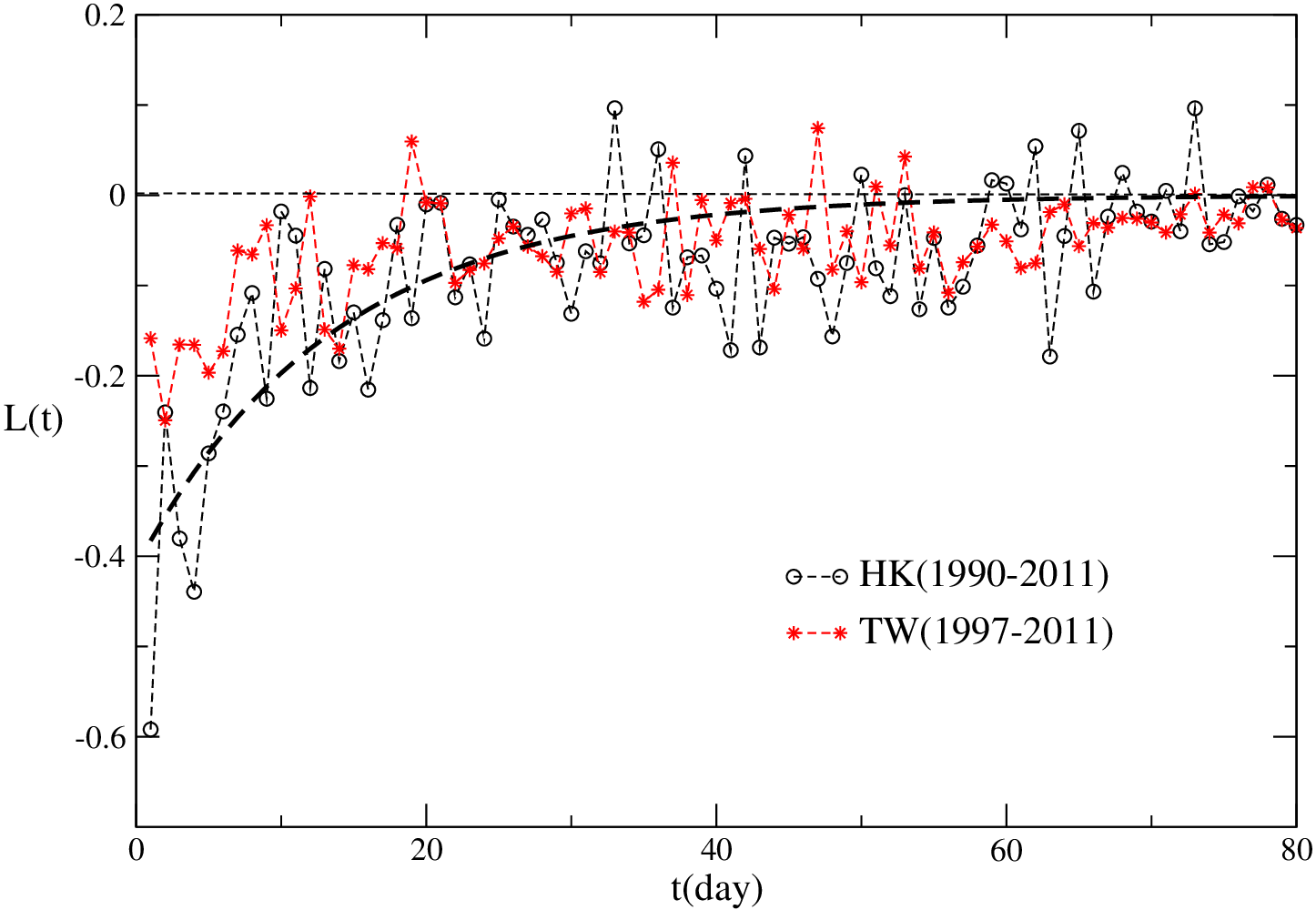}}{\footnotesize \par}

\caption{The return-volatility correlation of the Hang Seng Index (HK) from the year 1990 to 2011, and that of the Taiwan Weighted Index (TW) from 1997 to 2011. The dashed line shows an exponential fit.}

{\footnotesize \label{f8}}
\end{center}
\end{figure}

\begin{figure}[ht]
\begin{center}
\includegraphics[scale=0.35]{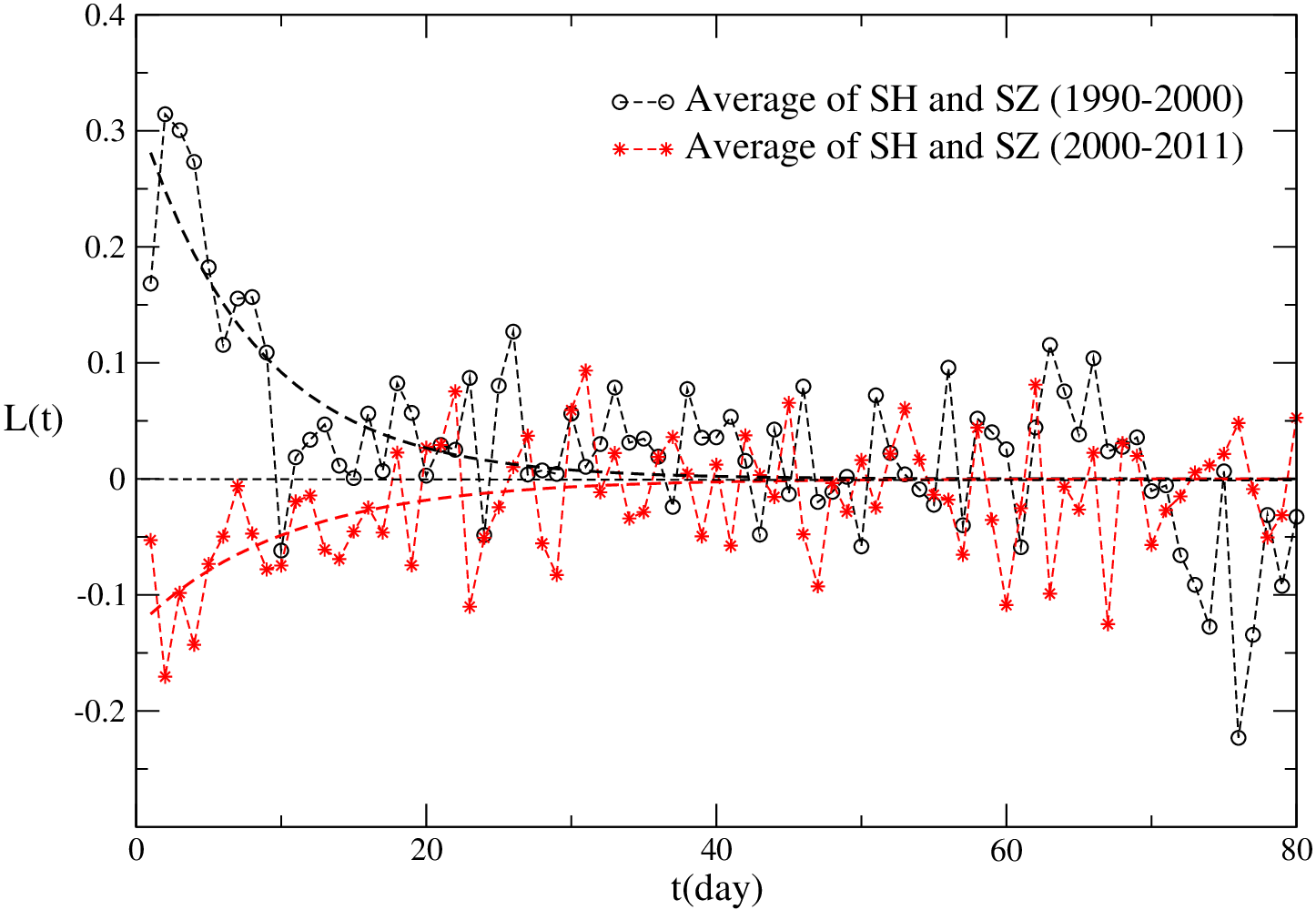}

\caption{$L(t)$ for the Shanghai Composite Index (SH) and Shenzhen Composite Index (SZ) in two time periods.
Dashed lines show the exponential fits.}

{\footnotesize \label{f10}}
\end{center}
\end{figure}

We compute the return-volatility correlation for the four stock markets in Greater China, with the daily price returns of each stock-market index. The results are displayed in Fig.\,\ref{f8} and Fig.\,\ref{f10}. $L(t)$ of the Hongkong stock market shows negative values up to at least 20 days. This is the well-known leverage effect. Similar to the Hongkong stock market, the Taiwan stock market also exhibits a leverage effect.
Qualitatively, $L(t)$ behaves the same for the Shanghai and Shenzhen stock markets, therefore we take the average over the two stock-market indices.
The removing of non-tradable shares was proposed around the year 2000. It is a significant event in the stock markets of mainland China. Therefore, we divide the whole period into two sub-periods by before and after the year 2000. We observe that before the year 2000, $L(t)$ shows positive values up to 10 days, i.e., an anti-leverage effect. After 2000, however, $L(t)$ displays negative values up to 20 days, i.e., a leverage effect.
This crossover behavior from the anti-leverage to the leverage effect is a new observation for the Shanghai and Shenzhen stock markets.
It seems that the two stock markets may be on the way gradually approaching a mature one.

One important reason for the crossover from the anti-leverage effect to the leverage effect in the Shanghai and Shenzhen stock markets may be
the removing of non-tradable shares around the year 2000.
In early years, the stocks were divided into tradable shares and non-tradable shares.
After that, all shares are tradable, which makes the two stock markets more regulated and mature.

On the other hand, the trading volume is an important variable which reflects the liquidity of the financial
markets.
To analyze the dynamic behavior of the trading volume, we introduce two basic measures: the volume return $R_{v}$ and the volume
volatility $\nu_{v}$, following the Refs.~\cite{pod09,li11,ren10}. The $R_{v}$ is defined as the logarithmic
change in the successive daily trading\begin{equation}
R_{v}(t)\equiv ln[V(t)/V(t-1)],\end{equation}
where $V(t)$ is the daily trading volume at time $t$. The normalized
return is\begin{equation}
r_{v}(t)=\frac{R_{v}(t)-\langle R_{v} \rangle}{\sqrt{\langle R_{v}^{2} \rangle -\langle R_{v} \rangle ^{2}}},\end{equation}
the volatility is the absolute value of return: $\nu_{v}=|r_{v}|$.

\begin{figure}[ht]
\begin{center}
{\footnotesize \includegraphics[scale=0.35]{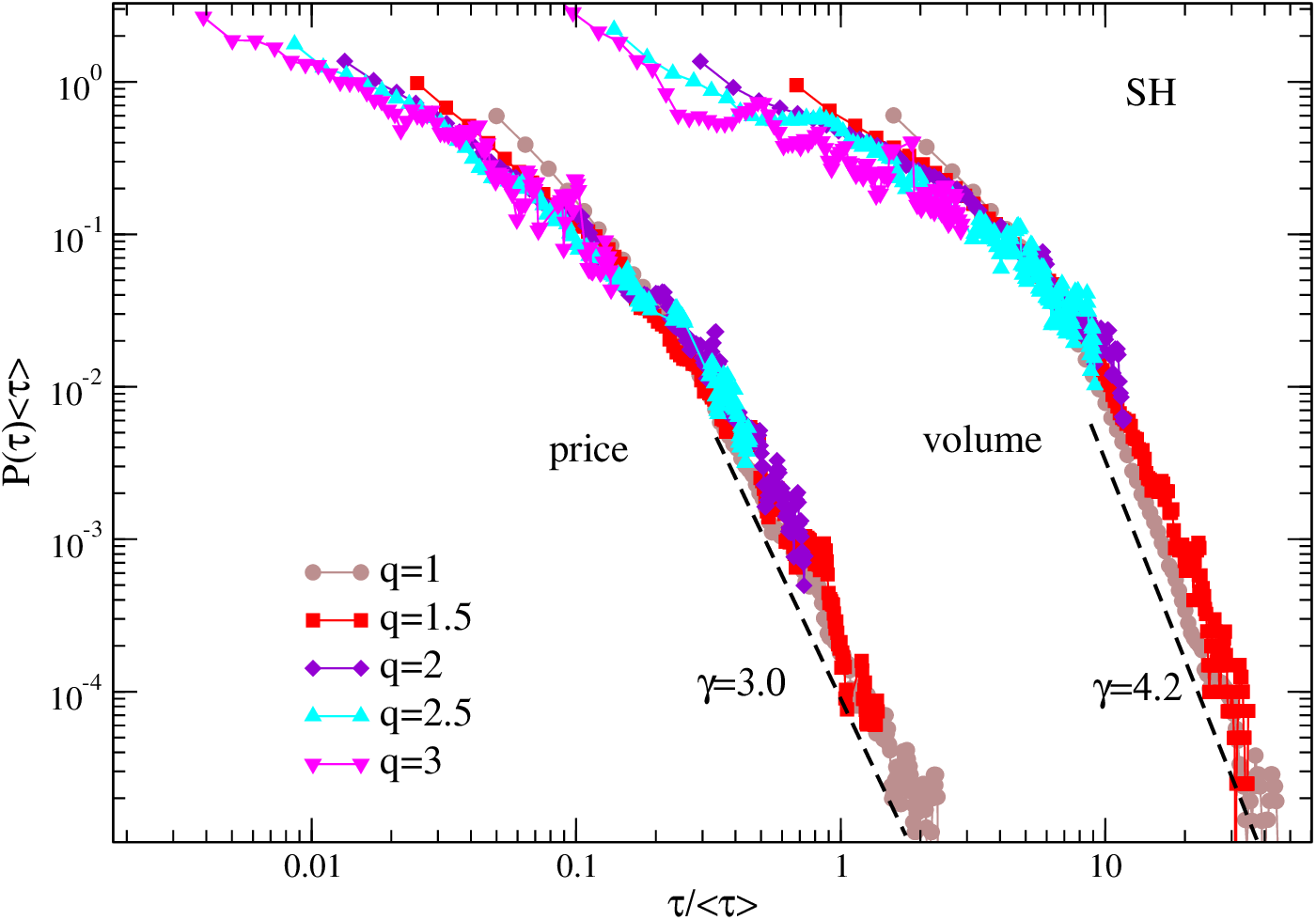}}{\footnotesize \par}

\caption{Probability distributions of the scaled recurrence intervals of the volume and price volatilities for different thresholds in the SH stock market. $\gamma$ is the power law exponent.}

{\footnotesize \label{f17}}
\end{center}
\end{figure}

From a volume volatility time series, we may extract the recurrence intervals $\tau$ between consecutive volatilities above a threshold $q$, and construct a series of the recurrence intervals
$\left\{ \tau(q)\right\} $~\cite{ren08,ren09,ren09a,qiu08}.
The threshold $q$ is measured in unit of the standard deviation of the volume volatility.
It can not be too large, otherwise we could not have sufficient data points.
Let us denote the probability distribution function of $\left\{ \tau(q)\right\}$ as $P_{q}(\tau)$.
The dependence of $P_{q}(\tau)$ on $q$ for the Shanghai stock market is shown in Fig.\,\ref{f17}.
After the probability of the recurrence intervals $P_{q}(\tau)$ being scaled with the mean interval $\langle \tau(q) \rangle$,
all the curves for different thresholds collapse onto a single curve, and it suggests that $P_{q}(\tau)$ obeys a scaling
function\begin{equation}
P_{q}(\tau)=\frac{1}{\langle \tau \rangle}f(\tau/\langle \tau \rangle).\end{equation}
As the threshold $q$ increases, the curve tends to be truncated due to the limited size of the data set~\cite{li11}. However, the tails of the scaling
function can be approximated by a power law \begin{equation}
f(\tau/\langle \tau \rangle)\sim(\tau/\langle \tau \rangle)^{-\gamma},\end{equation}
which is displayed by the dashed line in the figure, and $\gamma$ is the tail exponent.

We also investigate the statistical properties of the recurrence intervals constructed from the price volatility time series. The results for the Shanghai stock market are shown in Fig.\,\ref{f17}. The distributions diverge slightly for small intervals, but exhibit
scaling behaviors for large intervals. The method proposed in Ref.~\cite{pre11} is widely used to test the power-law fit~\cite{li11,ren10,ren08,ren09,ren09a,qiu08}. With this method, the power-law behavior of recurrence interval distributions for both price volatilities and volume volatilities pass the Kolmogorov-Smirnov test. In other words, our results support the ones in Ref.~\cite{li11} that the recurrence interval distribution of the price volatilities may follow the power-law behavior. The reason for the difference between our results and those in some previous papers~\cite{ren10,ren08,ren09,ren09a,qiu08} may be due to that our results are from the average over 259 stocks.

The values of the power law exponent $\gamma$ are shown in the last two columns of Table~\ref{t0}.
The exponents for the volume volatilities range from $3.7$ to $4.7$.
$\gamma=4.7$ for the Taiwan market is the largest one,
while, $\gamma=3.7$ for the Hongkong market is the smallest one. The exponents for the Shanghai and Shenzhen markets are $4.2$ and $4.3$ respectively.
However, the exponents for the price volatilities of the four stock markets are about $3.0$. The results indicate that the dynamic behavior of large price volatilities is rather robust, while that of large volume volatilities is not.

\section{Conclusion}

In the RMT analysis, after taking into account the signs of the components in an eigenvector of the cross-correlation matrix, one detects that a sector may split into two subsectors, which are anti-correlated with each other in the corresponding eigenmode.
For the four stock markets in greater China, the sector and subsector structures exhibit different characteristics.
The Shanghai and Shenzhen markets are dominated by the Basic materials and Industrial goods subsectors. For the Taiwan market, the dominating subsectors are Electronic industry and Chemical industry, while those for the Hongkong market are Real estate \& Finance and Service.
All these results reflect the features of the regional economies.

Meanwhile, we analyze the return-volatility correlation function. The Hongkong and Taiwan markets show a leverage effect.
However, the Shanghai and Shenzhen markets are more complicated. The two markets exhibit a strong anti-leverage effect before 2000, while it gradually changes to the leverage effect after 2000. This is a new observation for the stock markets in mainland China, additional to the discovery of the anti-leverage effect in Ref.~\cite{qiu06}.
We also study the recurrence interval distributions, and find that the power law exponents for the volume volatilities range from $3.0$ to $5.0$ for the four markets, while those for the price volatilities are about $3.0$.

\bibliographystyle{elsarticle-num}
\bibliography{eco2,zheng}

\newpage

\begin{table}
\begin{center}
\begin{tabular}{c|c|ccccc|cc}
\hline
 & Time period  & $T$ & $\lambda_{min}^{ran}$ & $\lambda_{max}^{ran}$ & $\lambda_{min}^{real}$ & $\lambda_{max}^{real}$ & $\gamma_{p}$ & $\gamma_{v}$\tabularnewline
\hline
SH & 2003.1-2011.7 & 2067 & 0.42 & 1.83 & 0.02 & 98.0 & 3.0 & 4.2\tabularnewline
SZ & 2003.1-2011.4 & 2000  & 0.41 & 1.85 & 0.12 & 98.0 & 3.1 & 4.3\tabularnewline
TW & 2003.1-2011.11 & 2206 & 0.43 & 1.80 & 0.14 & 72.3 & 3.2 & 4.7\tabularnewline
HK & 2003.1-2011.9 & 2146 & 0.43 & 1.82 & 0.12 & 35.5 & 3.2 & 3.7\tabularnewline
\hline
\end{tabular}
\caption{The second column shows the time periods of 259 stocks for the Shanghai (SH), Shenzhen (SZ), Taiwan (TW) and Hongkong (HK) stock markets. $T$ is
the total number of the daily data. $\lambda_{min(max)}^{ran}$ represents
the low (up) bound of the eigenvalues of the Wishart matrix, while $\lambda_{min(max)}^{real}$ is that of the real cross-correlation matrix. $\gamma_{p}$ is the power law exponent for the price volatility, and $\gamma_{v}$ is the one for the volume volatility.}
\label{t0}
\end{center}
\end{table}

\begin{table}
\begin{tabular}{c|cc|cc|cc|cc}
\hline
$\lambda_{i}$ & \multicolumn{2}{c|}{$\lambda_{1}$} & \multicolumn{2}{c|}{$\lambda_{2}$} & \multicolumn{2}{c|}{$\lambda_{3}$} & \multicolumn{2}{c}{$\lambda_{4}$}\tabularnewline
\hline
Signs & $+$ & $-$ & $+$ & $-$ & $+$ & $-$ & $+$ & $-$\tabularnewline
Sector & ST & IG & RE & Heal & Null & IG & BM & Ener\tabularnewline
$u_{c}$$=\pm0.08$ & $\nicefrac{23}{33}$ & $\nicefrac{11}{18}$ & $\nicefrac{19}{19}$ & $\nicefrac{10}{19}$ & $34$ & $\nicefrac{11}{17}$ & $\nicefrac{16}{26}$ & $\nicefrac{17}{28}$\tabularnewline
$u_{c}$$=\pm0.10$ & $\nicefrac{16}{25}$ & $\nicefrac{6}{6}$ & $\nicefrac{17}{17}$ & $\nicefrac{4}{7}$ & $18$ & $\nicefrac{5}{7}$ & $\nicefrac{7}{11}$ & $\nicefrac{16}{17}$\tabularnewline
\hline
\end{tabular}

\begin{tabular}{c|cc|cc|cc|cc>{\centering}p{10mm}>{\centering}p{10mm}>{\centering}p{10mm}>{\centering}p{10mm}}
\hline
$\lambda_{i}$ & \multicolumn{2}{c|}{$\lambda_{5}$} & \multicolumn{2}{c|}{$\lambda_{6}$} & \multicolumn{2}{c|}{$\lambda_{7}$} & \multicolumn{2}{c}{$\lambda_{8}$}\tabularnewline
\hline
Signs & $+$ & $-$ & $+$ & $-$ & $+$ & $-$ & $+$ & $-$\tabularnewline
Sector & EI & BM & BM & RE & BM & DG & IG & Null\tabularnewline
$u_{c}=\pm0.08$ & $\nicefrac{12}{23}$ & $\nicefrac{11}{26}$ & $\nicefrac{17}{32}$ & $\nicefrac{11}{23}$ & $\nicefrac{11}{22}$ & $\nicefrac{12}{24}$ & $\nicefrac{11}{22}$ & $27$\tabularnewline
$u_{c}=\pm0.10$ & $\nicefrac{9}{15}$ & $\nicefrac{8}{20}$ & $\nicefrac{6}{11}$ & $\nicefrac{5}{10}$ & $\nicefrac{9}{16}$ & $\nicefrac{7}{9}$ & $\nicefrac{7}{15}$ & $12$\tabularnewline
\hline
\end{tabular}

\caption{The subsectors for the SH stock market. The fraction is the number of well-identified stocks over the total number of stocks in the subsector. The abbreviations of the business subsectors are as follows. ST: Specially Treated stocks; RE: Real estate;
BM: Basic materials; Ener: Energy; Heal: Health care; DG: Daily consumer goods;
IG: Industrial goods; EI: Electronic industry; Null: No obvious category. Those non-fractions are the total counts of stocks in the corresponding subsectors.}

\label{t1}
\end{table}

\begin{table}
\begin{center}
\begin{tabular}{cc|c|c|c}
\hline
 &  & SZ & TW & HK\tabularnewline
\cline{1-5}
$\lambda_{1}$ & + & Ener & Null & Null\tabularnewline
 & - & IG & EI & RE\&Fin\tabularnewline
\hline
$\lambda_{2}$ & + & RE & RE & IG\tabularnewline
 & - & Heal & CI & RE\&Fin\tabularnewline
\hline
$\lambda_{3}$ & + & RE & DG & Serv\tabularnewline
 & - & Ener & SI & RE\&Fin\tabularnewline
\hline
$\lambda_{4}$ & + & Null & SI & Null\tabularnewline
 & - & RE & Null & RE\&Fin\tabularnewline
\hline
$\lambda_{5}$ & + & RE & DG & RE\&Fin\tabularnewline
 & - & DG & CI & Null\tabularnewline
\hline
$\lambda_{6}$ & + & Null & CI & RE\&Fin\tabularnewline
 & - & Ener & EI & Null\tabularnewline
\hline
$\lambda_{7}$ & + & Ener & DG & Null\tabularnewline
 & - & Null & CI & RE\&Fin\tabularnewline
\hline
$\lambda_{8}$ & + & EI & CI & Null\tabularnewline
 & - & DG & DG & RE\&Fin\tabularnewline
\hline
\end{tabular}

\caption{The subsectors for the SZ, TW and HK stock markets. CI: Chemical industry; SI: Steel industry; Serv: Service;
RE\&Fin: Real estate \& Finance. The other abbreviations can be seen in the caption of Table~\ref {t1}.}

\label{t2}
\end{center}
\end{table}

\newpage{}

\newpage{}

\end{document}